# Strong light-matter coupling in two-dimensional atomic crystals

Xiaoze Liu[1], Tal Galfsky[1], Zheng Sun[1], Fengnian Xia[2], Erh-chen Lin[3], Yi-Hsien Lee[3], Stéphane Kéna-Cohen[4] and Vinod M. Menon[1,†]

[1]Department of Physics, Graduate Center and Queens College, City University of New York, New York, New York, USA

[2]Department of Electrical Engineering, Yale University, New Haven, Connecticut, USA

[3]Department of Materials Science and Engineering, National Tsing Hua University, Hsinchu, Taiwan

[4]Department of Engineering Physics, École Polytechnique de Montréal, Montréal, Quebec, Canada

**Two dimensional (2D) atomic crystals of graphene, and transition metal dichalcogenides have emerged as a class of materials that show strong light-matter interaction. This interaction can be further controlled by embedding such materials into optical microcavities. When the interaction is engineered to be stronger than the dissipation of light and matter entities, one approaches the *strong coupling* regime resulting in the formation of half-light half-matter bosonic quasiparticles called microcavity polaritons. Here we report the evidence of strong light-matter coupling and formation of microcavity polaritons in a two dimensional atomic crystal of molybdenum disulphide (MoS$_2$) embedded inside a dielectric microcavity at room temperature. A Rabi splitting of 46 meV and highly directional emission is observed from the MoS$_2$ microcavity owing to the coupling between the 2D excitons and the cavity photons. Realizing strong coupling effects at room temperature in a disorder free potential landscape is central to the development of practical polaritonic circuits and switches.**

Two-dimensional (2D) materials have garnered much attention in the past decade owing to the plethora of exceptional electronic, mechanical, optical and thermal properties demonstrated by them. Specifically in the context of optoelectronics, the huge light-matter interaction that 2D materials have demonstrated has made them highly attractive for practical device applications. Graphene, the archetype 2D material was explored extensively for a wide array of photonic applications[1]. However, due to the lack of direct bandgap in graphene, considerable attention has shifted towards 2D materials known as layered transition metal dichalcogenides (TMDs)[2,3]. These TMDs are a group of naturally abundant material with a $MX_2$ stoichiometry, where M is a transition metal element from group VI (M = Mo, W); and X is a chalcogen (M = S, Se, Te). One of the most intriguing aspect of TMDs is the emergence of fundamentally distinct electronic and optoelectronic properties as the material transitions from bulk to the 2D limit (monolayer) [4–8]. For example, the TMDs evolve from indirect to direct bandgap semiconductors spanning the energy range of 1.1 to 1.9 eV in the 2D limit[4–6].

Among the TMDs, molybdenum disulphide ($MoS_2$) is one of the most widely studied systems used to demonstrate 2D light emitters[9], transistors[10,11], valleytronics[12–15] and photodetectors[16,17]. The novel excitonic properties of 2D $MoS_2$ that make it very interesting for both fundamental studies and applications include: (i) the enhanced direct band gap photoluminescence (PL) quantum yield of the monolayer compared with the bulk counterpart[7,8], (ii) the small effective exciton Bohr radius (0.93 nm) and associated large exciton binding energy (0.897 eV)[18,19] providing the opportunity for excitonic devices that operate at room temperature (RT) and (iii) the 2D nature of the dipole orientation making the excitonic emission highly anisotropic[20].

The interaction of a dipole with light can be modified by altering the surrounding dielectric environment. The most widely studied and technologically relevant phenomenon in this context is the Purcell enhancement wherein the spontaneous emission rate of the dipole is enhanced using an optical cavity by altering the photon density of states. Here the coupling between the dipole and the cavity photon is defined to be in the weak coupling regime since the interaction strength is weaker than the dissipation rates of the dipole and the photon. This regime has been demonstrated in the 2D materials using photonic crystal cavities coupled to 2D layers of $MoS_2$[21], and $WSe_2$[22]. It resulted in an enhancement of the spontaneous emission rate and highly directional photon emission.

When the interaction between the dipole and the cavity photons occur at a rate that is faster than the average dissipation rates of the cavity photon and dipole, one enters the strong coupling regime resulting in the formation of new eigenstates that are half-light – half-matter bosonic quasiparticles called cavity polaritons. Since with the pioneering work of Weisbuch *et al.*[23] there have been numerous demonstrations of cavity polariton formation and associated exotic phenomena in solid state systems using microcavities and quantum wells that support quasi 2D excitons[24–26]. However owing to the small binding energy of excitons in traditional inorganic semiconductors such as GaAs, most of these effects were observed at cryogenic temperatures making it impractical for real device applications. This issue has been partly mitigated using organic materials[27] and wide band gap semiconductors such as GaN and ZnO[28–30]. These systems, however, exhibit strong localization effects owing to the disordered potential landscape[31]. In this context 2D materials provide an ideal platform to realize polaritonic phenomena at room temperature in a mostly disorder free landscape. Here, we report for the first time the formation of strongly coupled 2D exciton- polaritons using a $MoS_2$ monolayer embedded in a dielectric microcavity (MC).

As shown in Fig. 1(a), the MC consists of a monolayer of $MoS_2$ sandwiched between two $SiO_2$ layers acting as the cavity layer which is placed between $SiO_2/Si_3N_4$ distributed Bragg reflector (DBR) mirrors. The $SiO_2$ layers and DBRs are fabricated by plasma enhanced chemical vapor deposition. The semiconducting monolayer $MoS_2$ is a thermodynamically stable form with a trigonal prismatic (2H-$MoS_2$) phase, where each molybdenum (Mo) atom is prismatically coordinated by six surrounding sulfur (S) atoms[4,32]. Large-area $MoS_2$ atomic layers (SEM image in Fig. 1(a)) are synthesized by chemical vapor deposition (CVD). The CVD $MoS_2$ exhibits a crystalline structure on various amorphous surfaces[33,34], and can be efficiently transferred with large areas onto the MC structure via aqueous solution transfer[34]. Details of sample fabrication are discussed in the Methods Section.

To determine the optical quality of the CVD grown $MoS_2$ used in the MC experiments reflectivity and steady state PL measurements are first carried out on as grown $MoS_2$ samples. Results of these measurements are shown in Supplementary Figure S1. The differential reflectivity ($\Delta R/R$), clearly shows two prominent absorption peaks at 1.872 eV and 2.006 eV, identified as A and B excitons, respectively[7,8] and the PL spectrum shows one dominant emission peak at 1.872 eV resulting from the direct bandgap transition of exciton A ($ex_A$)[7,8,32–34]. These results are consistent with previous reports of CVD grown monolayer $MoS_2$[33,34] indicating the presence of high quality monolayers.

Shown in Fig. 1(b) is the reflectivity of the MC structure embedded with $MoS_2$ at two different locations. As the $MoS_2$ monolayers are transferred to the MC, the existences of $MoS_2$ flakes at certain areas define the "active" and "passive" modes for this sample. The reflectivity curve labelled as "passive" is from a region where there is no MoS$_2$ layer, thus giving the bare cavity response. The cavity resonance is at 1.875 eV with a full width half maximum (FWHM) of $\hbar\Gamma_{cav} = 17$ meV. The cavity resonance is designed to overlap with the ex$_A$ energy. The "active" curve corresponds to a region where there is a monolayer of MoS$_2$ present in the cavity. The monolayer character of this region was verified from absorption and PL measurements as shown previously[4–7,32–34]. The reflectivity spectrum from the "active" region shows two distinct dips energetically shifted from the bare exciton and the cavity resonance indicating the presence of new eigenmodes.

To confirm the formation of strongly coupled polariton states in the active spot of the MC, angle-resolved reflectivity measurements are carried out. Reflectivity spectra for TM polarization at various incidence angles ranging from 7.5° to 30° are shown in Fig. 2. The spectra for TE polarization (not shown here) are almost identical. At small angles ($\leq 20°$), two prominent modes are observed, identified as the lower polariton branch (LPB) and the upper polariton branch (UPB). For clarification, the spectrum at 7.5° and 20° angles of incidence are expanded and shown in the inset of Fig. 2. Two pronounced modes corresponding to the LPB and UPB can be observed. The LPB blue-shifts with increasing angle and approaches ex$_A$ while UPB shifts away from ex$_A$ with increasing angle. At large angle ($> 20°$), the LPB slowly vanishes and only UPB is visible. The red solid lines trace the dispersion of both branches, indicating a clear anti-crossing feature around 20°. The experimental dispersion extracted from the measured reflectivity data, as shown in Fig. 3(a) is fit to a two coupled oscillator model (solid blue lines) showing a Rabi splitting of $\hbar\Omega_{Rabi} = 46$ meV (see Methods section). Based on the linewidths of exciton and cavity photon ($\hbar\Gamma_{ex} = 60$ meV and $\hbar\Gamma_{cav} = 17$ meV), we can obtain the interaction potential $V_A$ = 31.5 meV. As the criterion in the strong coupling regime where $V_A^2 \gg (\frac{\hbar\Gamma_{ex} - \hbar\Gamma_{cav}}{2})^2$, here $V_A^2 = 991$ meV$^2$ is much greater than $V_A^2 \gg (\frac{\hbar\Gamma_{ex} - \hbar\Gamma_{cav}}{2})^2 = 462$ meV$^2$, confirming the strong coupling regime is achieved in this MC. Based on the coupled oscillator model, contributions from excitonic and photonic components are also calculated for both branches as shown in Fig. 3(b) and 3(c). For small angle (<20°), the LPB is more photon-like and UPB is more exciton-like, and vice versa for larger angle (>20°), indicating the exciton-photon mixed composition of the LPB and UPB modes in the MoS$_2$ MC structure. We did not observe strong coupling between the cavity photons and ex$_B$ excitons due to the large negative cavity mode detuning with respect to its energy. In addition, Angle-resolved reflectivity is also simulated as shown in the contour plot in Figure S2, where the transfer matrix method (TMM) is employed and a Lorentzian oscillator is used to simulate the absorption of ex$_A$ based on experimental data (see Supplementary Materials).

Monolayer MoS$_2$ supports 2D excitons that have highly oriented in-plane dipoles and is an ideal candidate to develop polarization selective emitters and switches[20]. Angle-resolved PL measurements are carried out to investigate the modification of emission from the 2D excitons that are strongly coupled to the cavity photons. The Angle-resolved PL of Fig. 4 (a) shows similar dispersion to the reflectivity as the red traced curves. Shown in Fig. 4 (c) is the PL spectrum at 7.5° indicating the presence of two emission peaks corresponding to the LPB and UPB. The green and red curves are based on multiple Lorentzian peaks based curve fitting to locate the exact peak positions. The dispersion as the circles with error bars, extracted from the peak positions at various angles is shown in Fig. 4 (b). Using the same coupled oscillator model as in the reflectivity dispersion, the PL dispersion is fitted well with a consistent Rabi splitting of 46 meV.

Unlike traditional strongly coupled emission where the LPB emission is stronger[35], here we observe the UPB to exhibit the stronger emission at large angles. For monolayer MoS$_2$ the PL intensity reaches a maximum at zero in-plane wavenumber, due to the highly anisotropic orientation of the dipoles[20]. However, when placed in the MC, the PL peak intensity shows an antenna-like emission pattern, where it starts with a weak intensity at zero in-plane wavenumber, and reaches a maximum at some intermediate in-plane wavenumber. This is caused by the interaction between the in-plane component of the cavity photons and MoS$_2$ excitons,

To better understand the emission properties of the MoS$_2$ MC, the simulated emission from an in-plane oriented dipole representing the in-plane excitons embedded in the dielectric MC structure is carried out (see the Supplementary Figure S3). If overlaid with experimental dispersion, the simulated emission not only shows consistent dispersion of polariton branches, but also indicates a luminescence pattern consistent with the experiment where the emission intensity increases with angle up to 30$^o$.

In summary, we demonstrate for the first time strong coupling between 2D excitons and cavity photons in a monolayer MoS$_2$ based MC with a Rabi splitting of $\hbar\Omega_{Rabi}$ = 46 meV. Angle resolved reflectivity spectra show clear anticrossing behavior and the formation of the lower and upper polariton branches at room temperature. Angle-resolved PL shows two branch emission consistent with the reflectivity dispersion. However the emission intensity profile is distinct from other MC polariton demonstrations in quantum wells, wires and dots due to the highly anisotropic nature of the exciton orientation. The present demonstration of the formation of MC polaritons at room temperature in the 2D materials having large exciton binding energy and direct bandgap opens up the possibility of realizing practical plaritonic devices such as spin switches[36,37].

**Methods**

**Sample structure**. The microcavity sample consists of 8.5 periods of $SiO_2/Si_3N_4$ bottom distributed Bragg reflector (DBR), 7.5 periods of top DBR and a monolayer of $MoS_2$ sandwiched between two $SiO_2$ spacer layers. The $SiO_2$ layers and DBRs were fabricated by plasma enhanced chemical vapor deposition (PECVD) on a glass substrate using a combination of nitrous oxide, silane, and ammonia as the reactive gases. The monolayer $MoS_2$ was synthesized at 650 °C by ambient pressure chemical vapor deposition (APCVD) using perylene-3,4,9,10-tetracarboxylic acid tetra-potassium salt (PTAS) as the seed on $SiO_2$/Si substrate. Sulfur powder and molybdenum oxide ($MoO_3$) were used as the precursors for the synthesis. The $MoS_2$ monolayer was then immersed in DI water and lifted off from the substrate into water. A droplet of $MoS_2$ water solution was applied onto the spacer layer of $SiO_2$ above the bottom DBR and gently heated at 50 °C to form the new monolayer as the active cavity layer.

**Optical characterization.** Angle-resolved reflectivity measurements were carried out using a home built goniometer set up attached to a cryostat with an angular resolution of 1°. A tungsten lamp is used as the light source and a charge-coupled device-based fiber is coupled spectrometer to collect the optical signals in reflectivity measurements with a spectral resolution of 0.5nm. In photoluminescence (PL) measurements, an Ar-ion laser (488 nm) is used as the excitation source; an objective lens, a monochromator (Horiba iHR 320) and a CCD detector (Horiba Symphony liquid nitrogen cooled silicon CCD detector) combination is used for the detection. The spectral resolution here is of ~0.3nm. Angle-resolved PL measurements were carried out using the same goniometer as in the reflectivity measurements and device-based fibers to guide the light to the PL detection system.

**Theory.** The polariton energy dispersion can be interpreted by a coupled oscillator model as:

$$\begin{pmatrix} E_{cav}(\theta) + i\hbar\Gamma_{cav} - E & V_A \\ V_A & E_{ex} + i\hbar\Gamma_{ex} - E \end{pmatrix} \begin{pmatrix} \alpha \\ \beta \end{pmatrix} = 0.$$

Here the cavity mode was determined as $E_{cav}(\theta) = E_{ph} / \sqrt{1 - (\sin\theta / n_{eff})^2}$ with cutoff photon energy $E_{ph}$ and effective refractive index $n_{eff} = 1.48$ based on the fact the cavity layer has more than 99% fill fraction for $SiO_2$. $E_{ex}$ is the energy of ex$_A$. $E$ are the eigenvalues corresponding to the energies of polariton modes. $\alpha$ and $\beta$ construct the eigenvectors, representing the weighting coefficients of cavity photon, and ex$_A$ in each polariton state, where $\alpha^2 + \beta^2 = 1$. $V_A$ is the interaction potentials between photon and each of the excitons. The eigenvalues can be deduced as

$$E = \left(\frac{E_{ex} + E_{cav}}{2}\right) + i\left(\frac{\hbar\Gamma_{ex} + \hbar\Gamma_{cav}}{2}\right) \pm \sqrt{V_A^2 - \frac{1}{4}(\hbar\Gamma_{ex} - \hbar\Gamma_{cav})^2}$$

where $\hbar\Omega_{Rabi} = 2\sqrt{V_A^2 - \frac{1}{4}(\hbar\Gamma_{ex} - \hbar\Gamma_{cav})^2}$ as the Rabi splitting[38]. In the PL and reflectivity dispersion, the Rabi splitting is both fitted to be 46 meV.

## Acknowledgements

Work at CUNY is supported partially by the National Science Foundation through Grant No. DMR 1120923 and Army Research Office through Grant No. W911NF1310001. We thank the Ministry of Science and Technology of the Republic of China (103-2112-M-007-001-MY3) for partial support of this research.

## Author contributions

V. M. and F. X. initiated the project. X. L., V. M., and F. X. designed the experiments. X. L. fabricated the microcavity samples and carried out the experiments. X. L. collected and analyzed

Figure Legends:

Figure 1. (Color online) **Structures and optical properties of chemical vapor deposition (CVD) molybdenum disulfide ($MoS_2$) microcavity.** (a) Schematic of microcavity structure, chemical structure of $MoS_2$ monolayer and SEM image of the $MoS_2$ monolayer, the scale bar in the SEM is 10μm. (b) Reflectivity spectra for the passive spot corresponds to the area without $MoS_2$ monolayer (black), active spot correspond to area with it (blue). The vertical red dashed line represents $MoS_2$ $ex_A$ energy.

Figure 2. (Color online) **Angle-resolved reflectivity spectrum of the microcavity.** (a) Angle-resolved reflectivity spectrum at TM polarization from 7.5° to 30°. The vertical red dashed line represents $MoS_2$ $ex_A$ energy, the red curves trace the dispersion of microcavity polariton modes. Expanded views of reflectivity spectral features at 20° are shown in (b) and at 7.5° in (c). (b), Both zoomed spectra show clearly two polariton states at both sides of $ex_A$ energy.

Figure 3. (Color online) **Angle-resolved reflectivity spectrum and energy dispersion of the microcavity.** (a) Energy versus angle dispersion is extracted from the angle-resolved reflectivity spectra. The red spheres with error bars are the energy modes from the reflectivity spectra, the horizontal black dashed line represents the $ex_A$ energy, the black dashed curve represents the cavity modes and the two black solid curves correspond to theoretical fit of the polariton branches via a coupled-oscillator model. Hopfield coefficients for the microcavity polariton branches (UPB for (b) and LPB for (c)) calculated via the coupled-oscillator model. The Hopfield coefficients show the compositions of the polairtons. Here the black stars correspond to the coefficients of the cavity photons and the red spheres correspond to those of the excitons.

Figure 4. (Color online) **Angle-resolved PL spectra of the microcavity.** (a) Angle-resolved PL spectrum at TM polarization from 2.5° to 30°. The vertical red dashed line represents $MoS_2$ $ex_A$

energy, the red curves trace the dispersion of microcavity polariton modes. Expanded view of PL spectral features at 7.5° is shown in (c). A relatively weak LPB PL peak around 1.826 eV as well as a prominent UPB peak at 1.876 eV is observed. The PL spectrum is also fitted to multiple Lorentzian peaks to locate the exact PL peaks positions. (b) Energy versus angle dispersion, extracted from the angle-resolved PL spectra, agrees well with the reflectivity dispersion. The dispersion is also fitted to a coupled-oscillator model with the same Rabi splitting of 46 meV.

**Figure 1:**

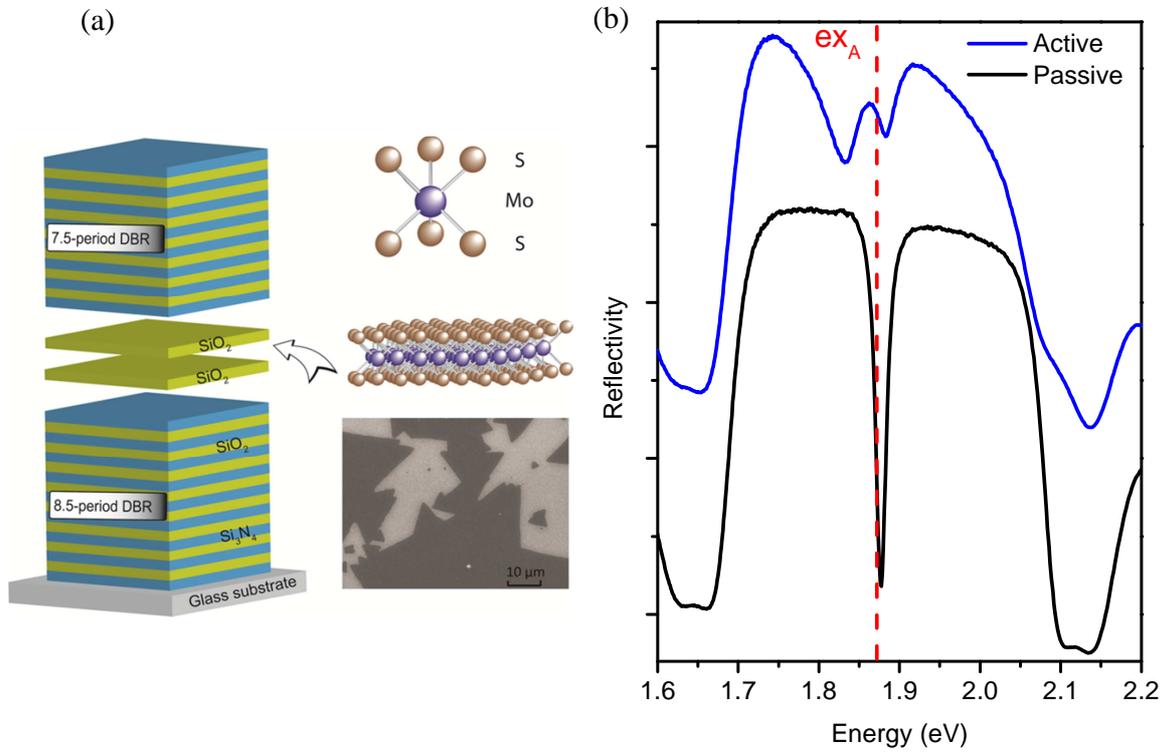

**Figure 2:**

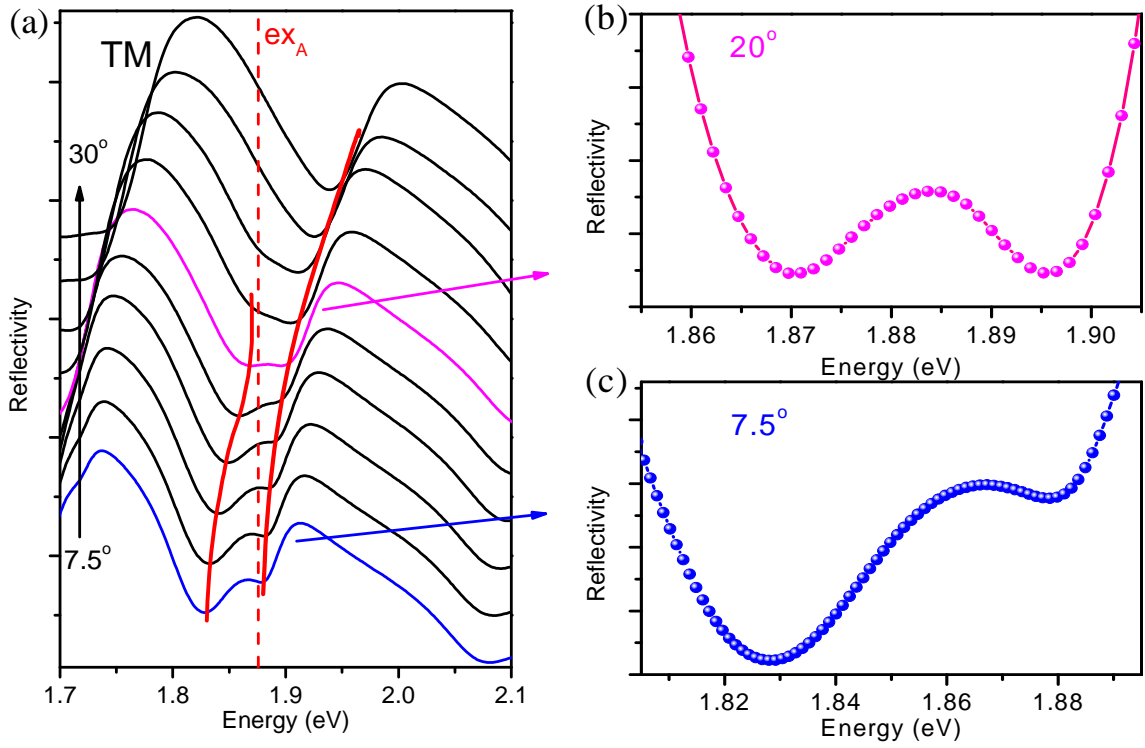

**Figure 3:**

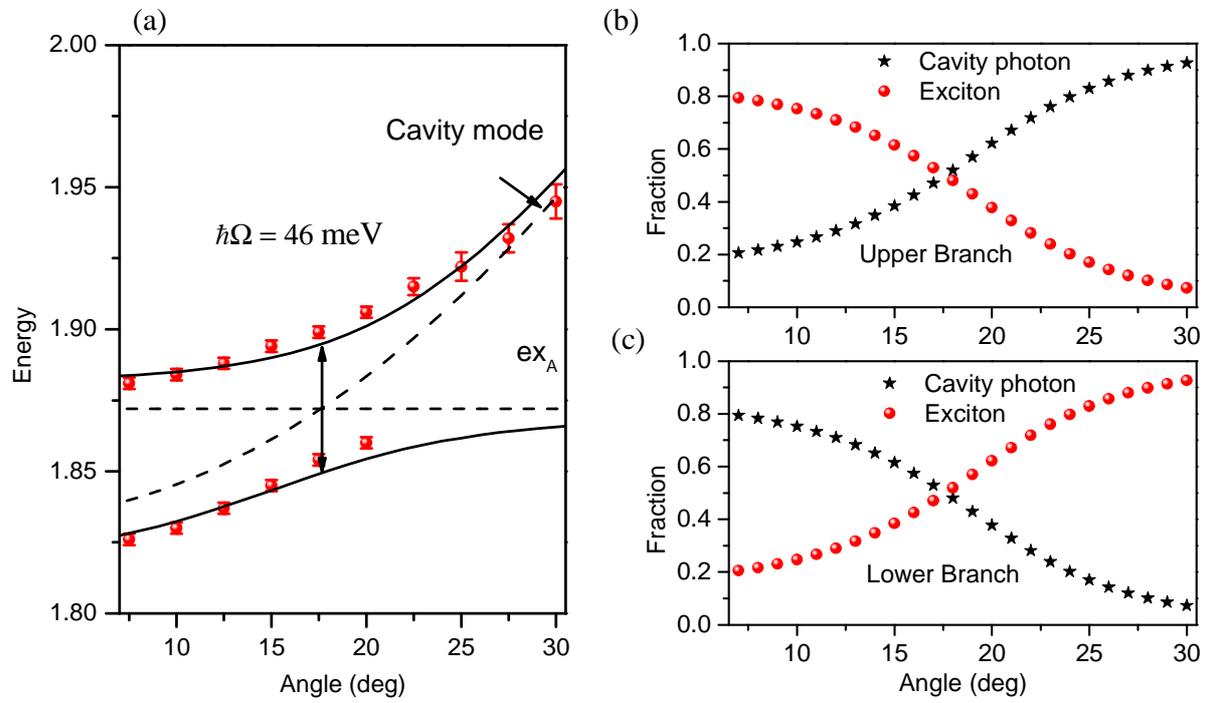

**Figure 4:**

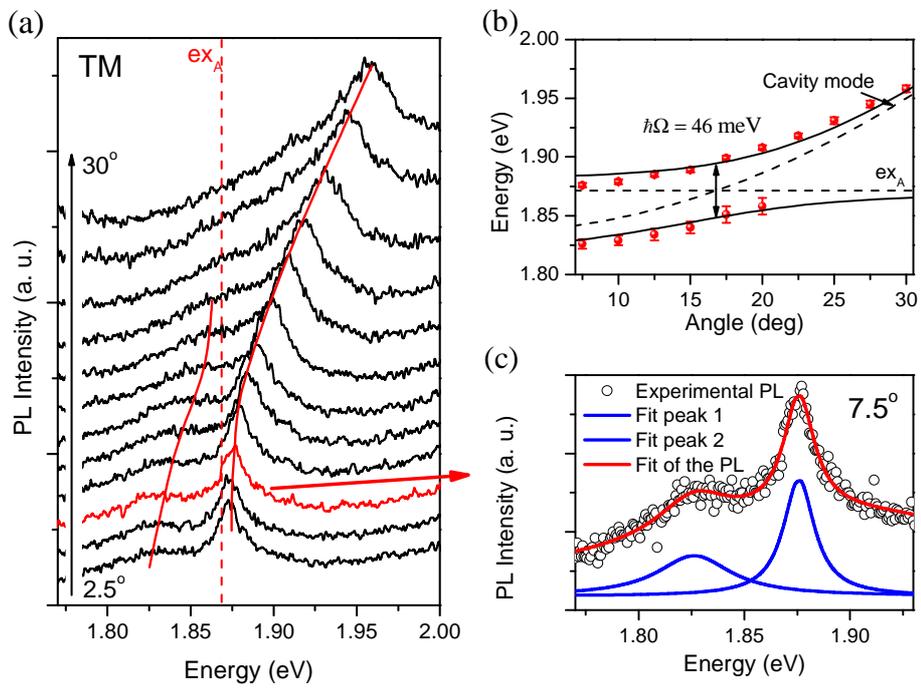

**Supplementary Material for:**

**Xiaoze Liu[1], Tal Galfsky[1], Zheng Sun[1], Fengnian Xia[2], Erh-chen Lin[3], Yi-Hsien Lee[3], Stéphane Kéna-Cohen[4] and Vinod M. Menon[1†]**

[1]Department of Physics, Graduate Center and Queens College, City University of New York, New York, New York, USA

[2]Department of Electrical Engineering, Yale University, New Haven, Connecticut, USA

[3]Department of Materials Science and Engineering, National Tsing Hua University, Hsinchu, Taiwan

[4]Department of Engineering Physics, École Polytechnique de Montréal, Montréal, Quebec, Canada

**Figure S1: Photoluminescence (PL) and differential reflectivity spectra of the MoS$_2$ monolayer.**

**Figure S2: Reflectivity spectrum at normal angle of the MoS$_2$ microcavity at various spots.**

**Figure S3: Simulated color map of angle-resolved PL for the MoS$_2$ microcavity**

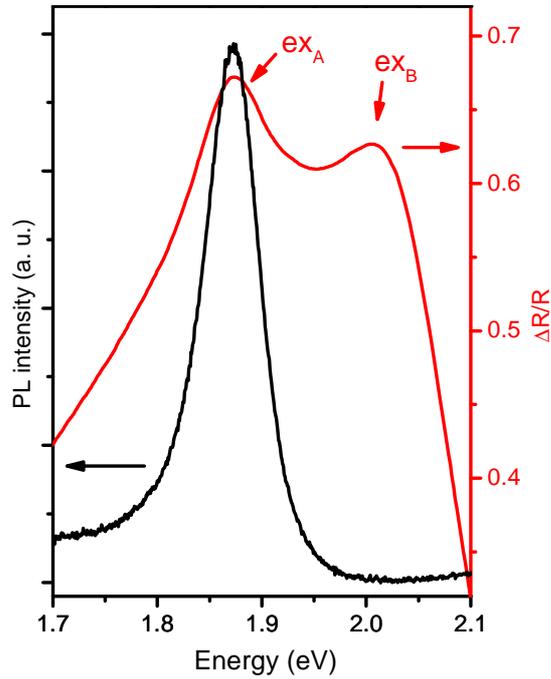

Figure S1. **Photoluminescence (PL) and differential reflectivity spectra of the MoS$_2$ monolayer.** The differential reflectivity spectrum of monolayer MoS$_2$ clearly shows two prominent absorption peaks at 1.872 eV and 2.006 eV, identified as A and B excitons, respectively[7]. The PL spectrum shows only one dominant peak at 1.872 eV resulting from the direct bandgap transition of exciton A (ex$_A$)[7]. These features of absorption and PL differentiates monolayer MoS$_2$ to bulk MoS$_2$ due to the transition of indirect bandgap to direct bandgap when it comes from bulk to monolayer[4,7].

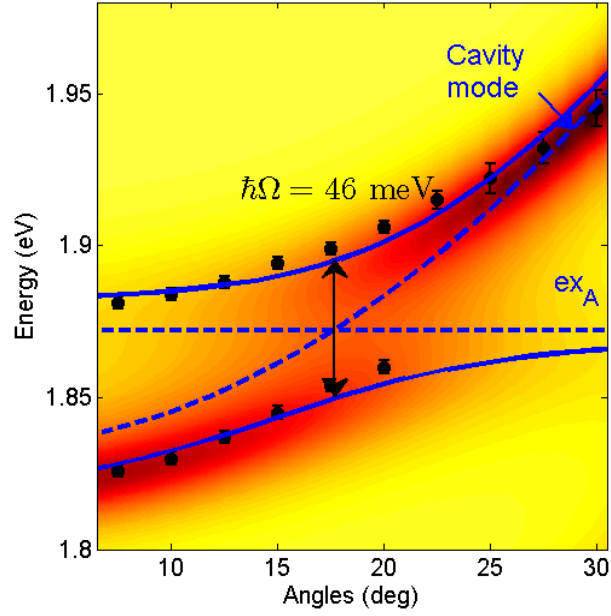

**Figure S2. Simulated angle-resolved reflectivity spectrum and energy dispersion of the microcavity.** Calculated reflectivity contour map for the MC, color gradient represents the reflectivity. Energy versus angle dispersion, extracted from the angle-resolved reflectivity spectra, overlays with contour map. The black spheres with error bars are the energy modes from the reflectivity spectra, the horizontal blue dashed line represents the $ex_A$ energy, the blue dashed curve represents the cavity modes and the two blue solid curves correspond to theoretical fit of the polariton branches via a coupled-oscillator model. The simulation shows good agreements with the measured data.

Given that we only observed coupling to $ex_A$ only, the dielectric function of monolayer $MoS_2$ is modeled by a Lorentzian oscillator:

$$\varepsilon(\omega) = \varepsilon_b + \frac{f}{\omega_0^2 - \omega^2 - i\Gamma_{ex}\omega}$$

Where $\hbar\omega_0$ is the ex$_A$ energy (1.87eV from the differential reflectivition spectrum in Figure S1), $\hbar\Gamma_{ex}$ is the linewidth of the exciton transition (60 meV from the PL spectrum), $\varepsilon_b$ is the background dielectric function, and $f$ is the oscillator strength. $\varepsilon_b$ and $f$ are adjustable parameters. With these assumptions, the predicted absorption coefficient $\alpha$ (at 1.87 eV) is $10^6$ cm$^{-1}$, which is consistent with reported data[7].

Refractive index of monolayer are derived from this model and applied to the transfer matrix method with a thickness of 0.75 nm[4,33] to simulate the MC angle-resolved reflectivity color map as in Figure 3(a).

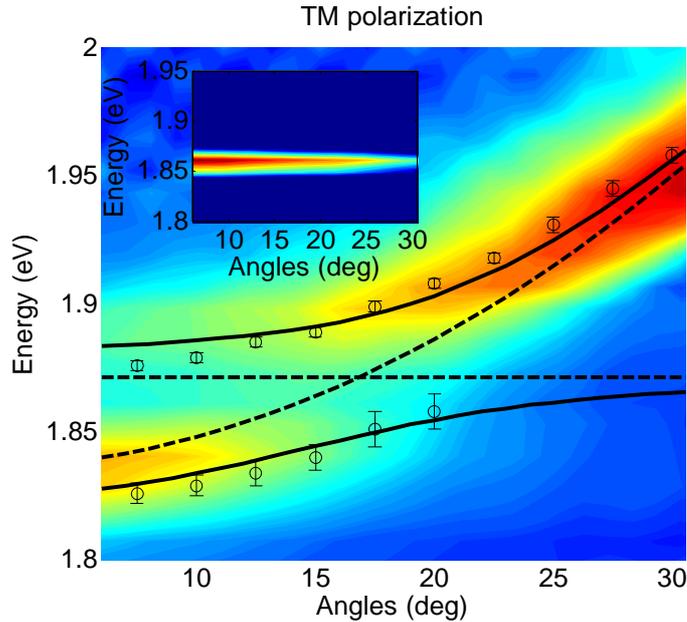

Figure S3. **Color map of angle-resolved PL for the MoS$_2$ microcavity.** Simulated color map for a horizontally aligned dipole emission from the MoS$_2$ microcavity showing modified emission dispersion as seen in the experiment. The black circles with error bars highlight the extracted experimental PL peak positions. For reference, the horizontal dashed line represents the ex$_A$ energy, the dashed curve represents the cavity modes and the two solid curves correspond to theoretical fit of the polariton modes. As expected, the PL intensity increases with the collection angle. Inset shows the emission pattern from the same horizontal dipole in the absence of the cavity where the MoS$_2$ is sandwiched between two SiO$_2$ layers.

The simulation is carried out using COMSOL for the microcavity structure with 0.75nm MoS$_2$ monolayer, which is in very good agreements with the experimental data. Interestingly, as in the inset, the monolayer MoS$_2$ PL, the intensity maximizes at zero in-plane wavenumber for TM polarization, due to the highly anisotropic orientation of the dipoles[20]. However, when placed in the MC, the PL peak intensity shows an antenna-like emission pattern, where it starts with a weak intensity at zero in-plane wavenumber, and maximizes at some intermediate in-plane wavenumber corresponding to an emission angle of ~30°. This is caused by the interaction between the in-plane component of the cavity photons and MoS$_2$ excitons, implying an efficient way to control the 2D exciton emissions via microcavity-exciton interaction.